\documentclass[a4paper]{article}
\usepackage[letterpaper, left=1in, right=1in, bottom=1in, top=0.75in]{geometry}
\usepackage{graphicx, xcolor}
\usepackage{amsfonts, amsmath,wasysym}
\usepackage{cancel}
\usepackage{scrextend}
\usepackage[numbers]{natbib}

\def\to{\rightarrow}

\def\R{\mathbb{R}}
\def\A{\mathbb{A}}

\def\({\left(} 	\def\){\right)}

\def\na{\nabla}
\def\Dal{\Box}
\def\then{\Rightarrow}
\def\union{\cup}

\def\al{\alpha}
\def\be{\beta}
\def\de{\delta}
\def\De{\Delta}
\def\si{\sigma}
\def\la{\lambda}
\def\ep{\epsilon}
\def\vp{\varphi}
\def\ga{\gamma}
\def\ka{\kappa}

\def\te{\theta}
\def\om{\omega}
\def\Ga{\Gamma}
\def\La{\Lambda}
\def\Om{\Omega}
\def\cR{{\mathcal{R}}}
\def\Frac[#1/#2]{\frac{#1}{#2}}
\def\dsi{\hbox{d}\si}

\def\Note{\medskip \begin{addmargin}[1cm]{1cm}\scriptsize}
\def\EndNote{\par\end{addmargin}\medskip\normalsize}
\def\<{\kern-1pt}

\renewcommand{\d}{\mathrm{d}}

\usepackage{graphics}
\begin{document}
\title{Solar System Tests in Brans-Dicke and Palatini $f(\cR)$-theories}
\author{A. Bonino$^{1}$, S. Camera$^{1,3}$, L. Fatibene$^{2,3}$, A. Orizzonte$^{2,3}$
%
%
\and
\small $^1$ Department of Physics, University of Torino, \\
\small via Pietro Giuria 1, 10125 Torino (Italy)
\and 
\small $^2$ Department of Mathematics, University of Torino, \\
\small via Carlo Alberto 10, 10123 Torino (Italy)
\and 
\small $^3$ Istituto Nazionale di Fisica Nucleare (INFN),  
\small Sez. di Torino (Italy),\\
\small via P. Giuria 1, 10125 Torino (Italy)
}
%
%
\maketitle
\abstract{
We compare Mercury's precession test in standard General Relativity (GR), Brans-Dicke theories (BD), and Palatini $f(\cR)$-theories.
We avoid post Newtonian (PN) approximation and compute exact precession  in these theories.
\\
We show that the well-known mathematical equivalence between Palatini $f(\cR)$-theories and a specific subset of BD theories does not extend to a really physical equivalence among theories 
since equivalent models still 
allow different incompatible precession for Mercury depending on the solution one chooses.
\\
As a result one cannot use BD equivalence to rule out Palatini $f(\cR)$-theories. 
On the contrary, we directly discuss that 
Palatini $f(\cR)$-theories can (and specific models do) easily pass Solar System tests as Mercury's precession.
} 
\section{Introduction}
Standard general relativity (GR) has proven to be a reliable theory of gravitation in many an instance (see also \cite{Virgo}). 
All predictions of the theory in vacuum have been confirmed up to the experimental precision, both locally in the Solar System as in astrophysical regimes such as in binary systems.

However, when considering non-vacuum solutions, as one does in galaxies and cosmology, standard GR is still successful,  yet at the price of introducing {\it dark sources} (dark matter and dark energy) to fit observations \citep[see][]{ClusterDM1,ClusterDM2,GalaxyDM1,GalaxyDM2,Lensing,DarkSectiorSupernovae}. While there are overwhelming pieces of evidence of the gravitational effects of such sources, there is still no direct evidence of their fundamental constituents. 

For these reasons, researchers have been considering the possibility that the effects which are currently ascribed to dark sources may be in fact purely gravitational effects due to modifications of the gravitational interaction itself. 
There are a number of candidate models which collectively are called {\it modified gravitational theories} or {\it extended theories of gravitation} \citep[see][]{MOND,CG1,CG2,CG3,Od1,Od2,Od3,ETC,Magnano}.
In these models there are no fundamental dark source fields or particles. It is the gravitational interaction that is modified with respect to standard GR,  and the modification is designed so that it preserves the local predictions in vacuum, while deviations at different scales justify observations in terms of {\it effective sources}. 

In this paper we consider two specific classes of modified models, Brans-Dicke models (BD) \citep{BD} and Palatini $f(\cR)$-theories
\citep[see][]{reviewPalatinif(R),Borowiec,Olmo2011,OlmoNostro}, especially in relation with classical Solar System tests, in particular Mercury's precession. The discussion aims also at highlighting that specifically in gravitational theories, observables are model dependent and when data are available, one needs to make predictions for each test within each extended model, making explicit all choices about observational protocols, since such choices may be different from what one does in standard GR.

The two modified models, Brans-Dicke and Palatini $f(\cR)$-theories, have been chosen because Palatini $f(\cR)$-theories are known to be dynamically equivalent via a conformal transformation, to a subset of Brans-Dicke models. Since Brans-Dicke theories have historically been used as a benchmark for Solar System tests, its parameters have been experimentally constrained and it has been shown that the preferred values of parameters correspond to standard GR. 

Now it happens that, by dynamical equivalence,  Palatini $f(\cR)$-theories correspond to a subset of Brans-Dicke theories (with a specific potential as well as) with a value of parameter which is not compatible with Solar System tests. Hence this equivalence has been used to rule out all Palatini $f(\cR)$-theories. 
This is not the only argument for ruling out Palatini $f(\cR)$-theories. It has also been argued (and confuted) that Palatini $f(\cR)$-theories
leads to singularities in polytropic stars \citep[see][]{no-go,Olmo,Mana,Wojnar}.

In this paper we shall show in details how this is wrong due to multiple reasons interacting with each other.  First, the dynamical equivalence requires a potential which was not assumed in the original Solar System tests analysis. Secondly, the value of the parameter for which the dynamical equivalence occurs is a singular value for Brans-Dicke models. Since the original analysis of Solar System tests in Brans-Dicke theories has been performed for generic regular values, one cannot even say that the singular value of the parameter has been ruled out (and in fact we shall show it is allowed). Lastly, we highlight how, even in view of the dynamical equivalence, test particles (e.g.\ Mercury itself) in the two theories are expected to go along different worldlines and in the specific example they do not only because one is considering a vacuum solution (so that the conformal factor is constant).

In general, we argue that dynamical equivalence may or may not extend to a complete physical equivalence, in which case the equivalence should preserve the action principle, as well as all the independent choices which define observational protocols \citep[see also][]{MathEquivalence}.

Test particles are an independent choice: when you use the eikonal approximation, equations for test particles are obtained, though they are not invariant with respect to redefinition of fields \citep[e.g.][]{ETG,Conservation}. Accordingly, the choice of test particle equations is just transformed into the choice of which field corresponds to test particle. Moreover, often one does not have a clear Lagrangian description of test particles in terms of fields and still one uses test particles.

Another choice, is space-time decomposition. In a relativistic theory there is no time and no space, just spacetime. Each observer may split spacetime into space-and-time, though each in a different way. Space and time lengths are thus relative to the choice and conventional. We use them extensively in astrophysics, just because a relativistic theory has no Dirac observables \citep{RovelliObs}. If we fix a space-time decomposition, that partially breaks general covariance, it conventionally reduces the symmetry group, so that non-trivial relative observables may be allowed.

One can show that defining atomic clocks then a space-time decomposition follows \citep[see][]{book2,Distances}, and one can define space and time lengths out of each specific atomic clock. That is very well known in standard GR, though it extends to a Weyl geometry \citep{Perlick}, as required in Palatini $f(\cR)$-theories.

In a previous paper (see \citep{Pinto1}; see also \citep{HubbleDrift}) we discussed cosmology in a particular Palatini $f(\cR)$-theory, based on the function
\begin{equation} 
f(\cR)= \al \cR -\Frac[\be/2]\cR^2 -\Frac[\ga/3]\cR^{-1}.
\label{fR}
\end{equation}
We discussed SNIa fit and showed that the system is quite strongly degenerate. If we provide $\al$ and $\be$, then the SNIa dataset allows to fix $\ga$,
by the way to a value of about $\ga\simeq 10^{-104}\,\mathrm{m^{-4}}$. That calls for independent measurements to fix $\al$ and $\be$.  Here we used Solar System tests to reduce degeneracy. It has also been argued (by an anonymous referee) that the best fit value $\al\simeq 0.1$ we found there would fail in Solar System tests, we show here that this is not the case considering Mercury test.

\medskip
Material is organised as follows.
In Section 2, we fix notation in BD theories and Palatini $f(\cR)$-theories.
In Section 3, we review the dynamical equivalence.
In Section 4, we consider static, spherically symmetric solutions which will be used to model the Solar System.
In Section 5, we consider geodesic equation, first for generic static, spherically symmetric metric, then for a solution.


\section{Brans-Dicke and  Palatini $f(\cR)$-theories}
A {\it Brans-Diche (BD) theory} is a gravitational theory for a metric $g$ and a scalar field $\vp$.
The action principle in dimension  $m=\dim(M)=4$ is
\begin{equation} 
L_{\rm BD}= \Frac[\sqrt{g}/2{\ka}] \(\vp R -\Frac[\om/\vp] \na_\mu \vp \na^\mu\vp+ U(\vp)\) \dsi\>,
\end{equation} 
where $R$ is the scalar curvature of $g$ and $U(\vp)$ is a potential.
The parameter $\om$ is called the {\it BD parameter}. 

From this action,  one has vacuum field equations
\begin{equation} 
\begin{cases}
\vp R_{\mu\nu} =  \na_{\mu\nu}\vp  
+ \Frac[\om/\vp] \na_\mu\vp\na_\nu \vp 
+ \Frac[1/2]  \( \Dal\vp
-U \) g_{\mu\nu} \\
\(3  + 2\om \) \Dal\vp 
+\(\vp U' -2 U  \)
=  0 \> . 	\\
\end{cases}
\end{equation} 
We shall eventually be interested also in the special case $\om=-\Frac[3/2]$
in which field equations become
\begin{equation} 
\begin{cases}
\vp R_{\mu\nu} =  \na_{\mu\nu}\vp  
- \Frac[3/2\vp] \na_\mu\vp\na_\nu \vp 
+ \Frac[1/2]  \( \Dal\vp
-U \) g_{\mu\nu} \\
\vp U' =2 U\>,  	\\
\end{cases}
\end{equation} 
which is of course a degenerate value since for $\om=-\Frac[3/2]$ the field equation for $\vp$ drops order and becomes an algebraic equation depending on the potential $U(\vp)$.
When no potential is assumed in the degenerate case $\om=-\Frac[3/2]$, the scalar field $\vp$ is left undetermined. The original analysis of Solar System tests \citep[see][]{Weinberg} was carried over with no potential and for a generic regular value of $\om$.

In BD models, test particles go along timelike geodesics of $g$, which determines the geometry of spacetime as well as its metric structure. The scalar field $\vp$ is non-minimally coupled and it modifies the law in which gravitational field is mediated.

 \subsection{Palatini $f(\cR)$-theory}
 
For a Palatini $f(\cR)$-theory, we start from fields $(g_{\mu\nu}, \tilde \Ga^\ep_{\mu\nu})$, where $\tilde \Ga$ is a (here torsionless) generic connection on the spacetime $M$, {\it a priori} independent of $g$, and a Lagrangian
\begin{equation} 
L_f= \Frac[\sqrt{g}/2{\ka}] f(\cR) \dsi 
\end{equation} 
for some (regular enough) function $f(\cR)$ of the scalar curvature $\cR = g^{\mu\nu} \tilde R_{\mu\nu}$, where $\tilde R_{\mu\nu}$ is the Ricci tensor of the connection $\tilde \Ga$ alone.
Field equations read
\begin{equation} 
\begin{cases}
 f^\prime(\cR) \tilde R_{(\mu\nu)} - \Frac[1/2] f(\cR)  g_{\mu\nu}=0\\
 \tilde \na_\epsilon \(\sqrt{g}f^\prime(\cR) g^{\mu\nu} \)=0\>.\\
\end{cases}
\end{equation} 
By tracing the first field equation by $g^{\mu\nu}$, we obtain the so-called {\it master equation},
\begin{equation} 
f^\prime(\cR) \cR - 2  f(\cR)  =0\>,
\end{equation} 
which must be identically satisfied along solutions. The function $f(\cR)$ is called {\it regular enough} when the zeros of the master equation are simple and they form a discrete set.

To solve the second equation, one has to set $\vp\propto f^\prime(\cR)$.
That can be locally inverted as $\cR\propto r(\vp)$.

\Note
For the model based on (\ref{fR}),
we have the master equation
\begin{equation} 
\al\cR -\cancel{\be\cR^2} + \Frac[\ga/3] \cR^{-1} -2\al\cR +\cancel{\be \cR^2} +\Frac[2\ga/3]\cR^{-1}
=\Frac[ \ga -\al\cR^2/ \cR] = 0
\quad\then
{}^\pm\cR= \pm \sqrt{\Frac[ \ga/\al ]}
\end{equation} 

With these values of the curvature we get
\begin{equation} 
f_\pm := f({}^\pm\cR) =\pm \Frac[2\sqrt{\al\ga} /3]  -\Frac[\be\ga/2\al] 
\end{equation} 
and
\begin{equation} 
\vp_\pm=f^\prime_\pm= f^\prime({}^\pm\cR) =   \Frac[ 4\al/3 ] \mp\be\sqrt{\Frac[ \ga/\al ]}
\end{equation}

At that point the (vacuum) field equations are
\begin{equation} 
\begin{cases}
R_{\mu\nu} =  \tilde R_{\mu\nu} =  \Frac[1/2\vp] f(\cR)  g_{\mu\nu} = \La_\pm g_{\mu\nu} 
\quad\then
\La_\pm= -\Frac[1/4] \Frac[{3\be\ga}   \mp {4\al\sqrt{\al\ga}  } / {4\al^2} \mp3\be\sqrt{\al \ga }]  
\simeq  \pm\Frac[1/4] \sqrt{\Frac[\ga/\al]} 
 \\
 f^\prime(\cR) \cR - 2  f(\cR)  =0
\quad\then
{}^\pm\cR= \pm \sqrt{\Frac[ \ga/\al ]}
\\
\end{cases}
\label{fReq}
\end{equation}
where the last approximations have been done for small values of $\be\ga$ with respect to $\al\sqrt{\al\ga}$ (which both have units of an inverse squared lenght)
and small values of $\be\sqrt{\al\ga}$ with respect to $\al^2$ (which are both adimensional)

\EndNote

Let us notice that, as a consequence of the assumption of being in vacuum, the conformal factor $\vp$ is constant 
and field equations reduce to Einstein equations with a comslogical constant, which is proven in general by universality theorem \citep[see][]{Universality}.
Universality theorem guarantees that vacuum solutions of Palatini $f(\cR)$-theories maintain the successes shown by standard GR solutions provided that the cosmological constant is small enough, as small as it is expected by observations.

In the specific example (\ref{fR}), the (effective) cosmological constant $\La_\pm$ is small enough iff $\ga$ is small enough
and $\ga$ is the parameter which is better constrained by SNIa. 
We can consider the best value of $\ga$ 
\begin{equation}
\ga \simeq  2.46^{+3.84}_{-2.24}\cdot 10^{-104} \,\mathrm{m}^{-4}  
\label{PintoPar}
\end{equation}
obtained in (\ref{fR}) setting $\al=0.095$, $\be=0.25 \,\mathrm{m}^{-2}$.

As a matter of fact, the predicted best fit value of $\La = 1.27^{+0.58}_{-0.98}\cdot 10^{-52}  \,\mathrm{m}^{-2}$ is not far away from the value observed, e.g.\ by the one found by Planck survey  ($\La_{\rm Planck (2018)} = (1.106 \pm 0.023) \cdot 10^{-52} \,\mathrm{m}^{-2}$) \cite[see][]{PlanckExp} 

This value for the (effective) cosmological constant will not be observable (or falsifiable) with experiments in the Solar System.
If it were, then $\La CDM$ would be falsified as well \citep[see][]{CMB,WMap}.

Thus let us review the dynamical equivalence and then investigate how this rough though clear result can be compatible with
exclusion of the values $\om< 4 \cdot 10^4$ \citep[see][]{Cassini1,Cassini2}, hence including $\om=-\Frac[3/2]$, by BD and Mercury's precession.

\section{Equivalence between Palatini $f(\cR)$-theories and BD}

Before going to field equations, one can prove dynamical equivalence at the level of the action.
The proof of equivalence is in two steps. In the first step we show equivalence between Palatini $f(\cR)$-theory and a theory with an extra scalar field governed by the Helmholtz Lagrangian.
In the second step we recast Helmholtz Lagrangian as a BD model by a suitable field transformation \citep[see][]{Magnano,ETG}.

Let us consider the definition of the conformal factor $\vp=f^\prime(\cR)$ and solve it for the curvature $\cR$,
\begin{equation} 
\vp= \al -\be \cR + \Frac[\ga/3] \cR^{-2}= \Frac[3\al \cR^2-3\be \cR^3 +\ga / 3\cR^2]
\quad\then
 \quad 
3\be \cR^3+ 3(\vp-\al)\cR^2 -\ga =0\>.
\end{equation} 

\Note
Let us fix  $\vp_\ast = \al + {}^3\sqrt{\Frac[3\be^2\ga/4]}$.
For $\vp< \vp_\ast$, the equations has only one positive solution $\cR={}^+\cR(\vp)$.

For $\vp\ge  \vp_\ast$. the equation has  one positive solution $\cR={}^+\cR(\vp)$ as well as 2 negative solutions $\cR={}^-\cR_1(\vp)$ and $\cR={}^-\cR_2(\vp)$.

\begin{figure}[htbp] 
   \centering
   \includegraphics[width=2in]{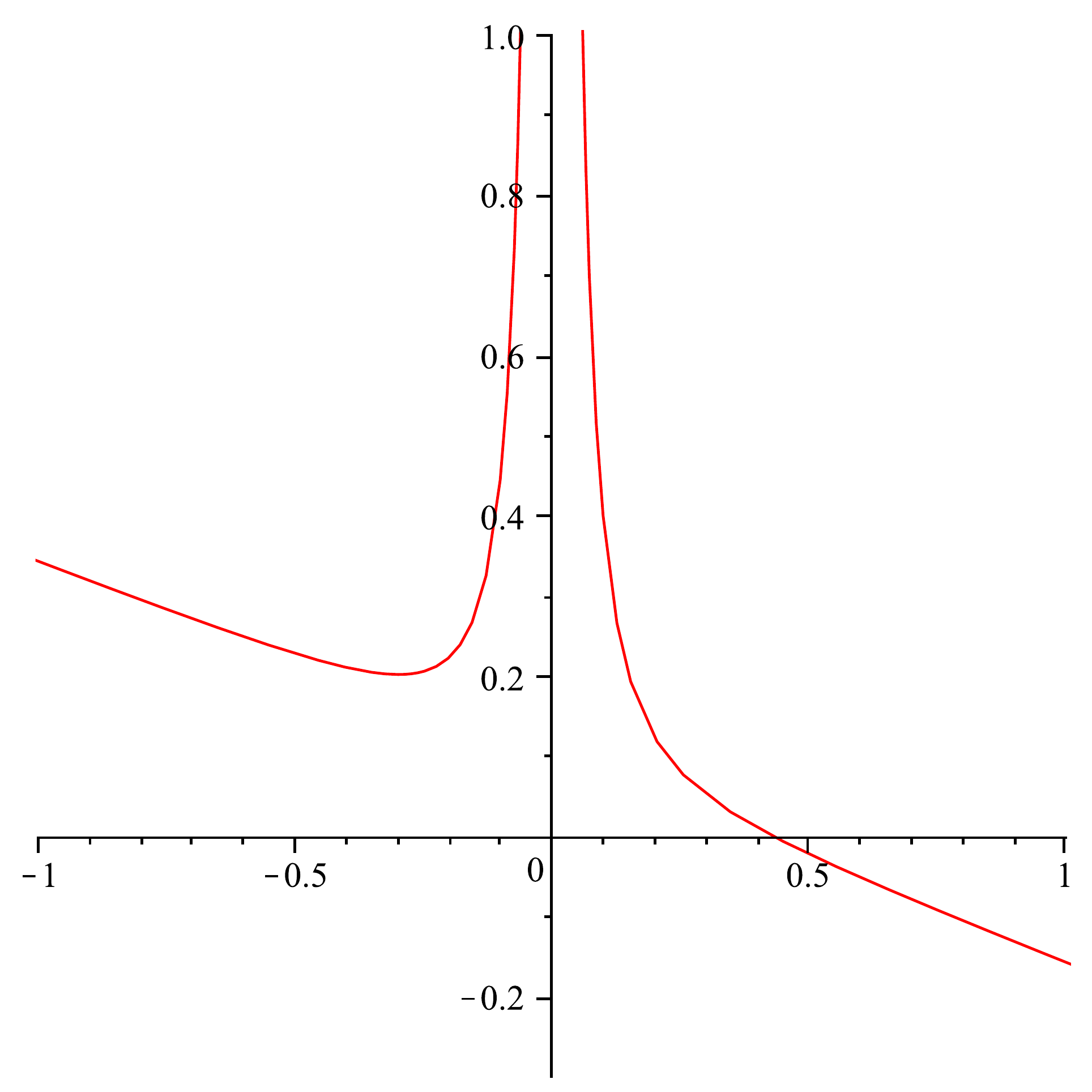} 
   \caption{\small \it $\vp(\cR)$ for $\al=0.091$, $\be=0.25 \,\mathrm{m}^{-2}$, $\ga=1/100 \,\mathrm{m}^{-4}$ in the plane $(\cR, \vp)$. The asymptotic line just depends on $\al$ and $\be$. The deviation from it is governed by $\ga$ (the smaller $\ga$,   the more the graph is closed to the asymptotic line). }
      \label{fig:example}
\end{figure}

\medskip

When one inverts for $\cR= r(\vp)$, there are 3 branches:

\begin{itemize}
\item[]i) $\cR>0$ and any $\vp\in \R$;  $\cR= r_1(\vp)$.

\item[]{ii)} $\cR_\ast\le\cR<0$ and $\vp\ge \vp_\ast$;  $\cR= r_2(\vp)$.

\item[]{iii)} $\cR\le\cR_\ast<0$ and $\vp\ge \vp_\ast$;  $\cR= r_3(\vp)$.
\end{itemize}

where we set $\cR_\ast = {}^-\cR(\vp_\ast)$.

The for each branch we have an `inverse' $\cR= r(\vp)$ and we can define a Helmholtz Lagrangian
\begin{equation} 
L_{\rm H}= \sqrt{g} \big( f(r(\vp))+ \(\cR- r(\vp)\)\vp\big) \dsi
\end{equation} 
which depends on $\vp$, $g$, $\tilde \Ga$ and first derivatives of $\tilde \Ga$.

\EndNote

The Helmholtz Lagrangians are dynamically equivalent to $f(\cR)$-theory.
By varying Helmholtz Lagrangian with respect to $(g, \tilde \Ga, \vp)$, one gets field equations
\begin{equation} 
\begin{cases}
\vp \tilde R_{(\mu\nu)} = \Frac[1/2] f(r(\vp)) g_{\mu\nu}\\
\tilde \na_\ep (\sqrt{g} \vp g^{\mu\nu})=0 \\
 \cR= r(\vp)\\
\end{cases}
\end{equation} 
the last equation being equivalent to a branch of $\vp= f^\prime(\cR)$.
Then we can define $\tilde g= \vp g$ and solve the second to get $\tilde \Ga= \{\tilde g\}$.
Finally, from the first one, one can trace to get the master equation (and get that $\cR$ and hence $\vp$ are constant on shell).
Since $\vp$ is constant $\{g\}=\{\tilde g\}$ and $\tilde R_{(\mu\nu)} = R_{\mu\nu} $.
Then, consequently, the first equation becomes Einstein with cosmological constant, equivalent to (\ref{fReq}).

As a matter of fact, this gives us the chance to test the Palatini $f(\cR)$ model directly without passing through BD equivalence.
Moreover, the other way around, this gives us also the chance to use equivalence to test the degenerate BD models which are equivalent to Palatini $f(\cR)$ models.
In both cases, we have that the value of the cosmological constant one gets from $f(\cR)$ given by (\ref{fR}), with parameters (\ref{PintoPar}),
is compatible with what found by Planck survey in 2018. 
Accordingly, the models are not rules out by solar system tests (as well as by other test which are insensitive to the observed value of the cosmological constant).

If we select a specific potential $U=  f(r(\vp))- \vp r(\vp) $ in a Brans-Dicke theory we get equations:
\begin{equation} 
\begin{cases}
\vp R_{\mu\nu} =  \na_{\mu\nu}\vp  
+ \Frac[\om/\vp] \na_\mu\vp\na_\nu \vp 
+ \Frac[1/2]  \( \Dal\vp
- f(r(\vp))+ \vp r(\vp)  \) g_{\mu\nu } \\
\(3  + 2\om \) \Dal\vp 
+\(\vp r(\vp) -2 f(r(\vp))  \)
=  0 	\>.\\
\end{cases}
\end{equation} 

\Note
With that potential, one has
\begin{equation} 
\begin{aligned}
\vp U' -2 U=& \vp\cdot (f^\prime(r(\vp)) r'(\vp) - r(\vp) - \vp r'(\vp) ) -2 f(r(\vp))+ 2 \vp r(\vp)=\\
=& \vp\cdot (\cancel{\vp r'(\vp)} - r(\vp) - \cancel{\vp r'(\vp)} ) -2 f(r(\vp))+ 2 \vp r(\vp)=\\
=& -\vp r(\vp)  -2 f(r(\vp))+ 2 \vp r(\vp)=     \vp r(\vp) -2 f(r(\vp))\\
\end{aligned}
\end{equation} 
which in fact is the left hand side of the master equation.

Ricci tensor conformal transformations are given by
\begin{equation} 
\vp\tilde R_{\mu\nu} +\na_{\mu\nu} \vp+\Frac[1/2]\Dal\vp g_{\mu\nu} - \Frac[3/2\vp] \na_\mu\vp \na_\nu\vp = \vp R_{\mu\nu} \>.
\end{equation} 
Thus the first equation reads as
\begin{equation} 
\vp\tilde R_{\mu\nu} +\cancel{\na_{\mu\nu} \vp}+\bcancel{\Frac[1/2]\Dal\vp g_{\mu\nu}} - \Frac[3/2\vp] \na_\mu\vp \na_\nu\vp = 
\cancel{ \na_{\mu\nu}\vp  }
+ \Frac[\om/\vp] \na_\mu\vp\na_\nu \vp 
+ \Frac[1/2]  \( \bcancel{\Dal\vp}
- f(r(\vp))+ \vp r(\vp)  \) g_{\mu\nu} 
\end{equation} 
\begin{equation} 
\vp\tilde R_{\mu\nu}  = 
 \Frac[1/2\vp] \(3+2\om   \)\na_\mu\vp\na_\nu \vp 
 +  \Frac[1/2]  \(- f(r(\vp))+ \vp r(\vp)  \) g_{\mu\nu} \> .
\end{equation} 

By tracing the first equation, we get
\begin{equation} 
\vp \cR =   \Frac[2\om+3/2\vp] \na_\ep\vp\na^\ep\vp 
+ 2 \vp \cR - 2f(r(\vp))  
\quad\then 
\vp \cR - 2f(\cR)   = -  \Frac[2\om+3/2\vp] \na_\ep\vp\na^\ep \vp\> . 
\end{equation}

\EndNote

For $\om\not=-\Frac[3/2]$
\begin{equation} 
\begin{cases}
\vp\tilde R_{\mu\nu}  = 
 \Frac[3+2\om/2\vp] \na_\mu\vp\na_\nu \vp 
 +  \Frac[1/2]  \(\vp r(\vp)  - f(r(\vp)) \) g_{\mu\nu} =\\
\quad\quad
 =  \Frac[3+2\om/2\vp] \na_\mu\vp\na_\nu \vp 
 +  \Frac[1/2]  \(\vp r(\vp)  - 2f(r(\vp)) \) g_{\mu\nu} + \Frac[1/2]f(r(\vp))  g_{\mu\nu} \\
\quad\quad
 =  \Frac[3+2\om/2\vp] \(\na_\mu\vp\na_\nu \vp 
 + \vp \Dal \vp g_{\mu\nu} \) + \Frac[1/2]f(r(\vp))  g_{\mu\nu} 
 \\
\Dal\vp =\Frac[ \vp r(\vp) - 2f(r(\vp)) / (3+2\om)]\>.
\end{cases}
\end{equation} 
There are solutions of the second equation with a non-constant $\vp$.

For  $\om=-\Frac[3/2]$
\begin{equation} 
\begin{cases}
\vp\tilde R_{\mu\nu}  = 
  \Frac[1/2]  \(- f(r(\vp))+ \vp r(\vp)  \) g_{\mu\nu} =  \Frac[1/2] f(r(\vp))g_{\mu\nu}
 \\
\vp f(r(\vp)) - 2f(\cR)    =0	\> .
\end{cases}
\end{equation} 

The second equation implies that $\vp$ is constant, then $\tilde R_{\mu\nu}= R_{\mu\nu}$.
By tracing the first, one gets the master equation, from which $\cR$ is constant.
Then the first equation becomes Einstein with cosmological constant.

That shows that there is a dynamical equivalence between Palatini $f(\cR)$-theories and 
BD theories with $\om=-\Frac[3/2]$ and a potential $U(\vp)= f(r(\vp))- \vp r(\vp)$.

For the function $f(\cR)$ given by (\ref{fR}), the function $r(\vp)$ is quite complicated, but in fact we do not really need it.
Anyway, currently we are on the positive branch (the first one) and as long as observational cosmology is concerned we can restrict to that branch.

For a proof at the level of action see \cite{Magnano,ETG}.

\section{Solutions with point-like sources}

Let us here consider the static spherically symmetric solutions in Palatini $f(\cR)$-theory and BD theory, both in the case of generic parameter
and no potential and for $\om=-\Frac[3/2]$ and the potential $U(\vp)= f(r(\vp))- \vp r(\vp)$ induced by dynamical equivalence.

\subsection{Solution in Palatini $f(\cR)$-theory}

In a Palatini $f(\cR)$-theory, in view of universality theorem, we get a static spherically symmetric solution which is
\begin{equation} 
\tilde g = -A(r) \d t^2 + \Frac[ \d r^2/ A(r)] + r^2 \d\Om^2
\quad\textrm{with}\quad
A(r)  = a  - \Frac[b/r] - \Frac[\La/3] r^2\>,
\end{equation} 
where we set $\d\Om^2 := \d\te^2 + \sin^2(\te) \d\phi^2$ for the volume element on the sphere. Thus 
\begin{equation} 
g= \vp^{-1} \(-A(r) \d t^2 + \Frac[ \d r^2/ A(r)] + r^2 \d\Om^2\)\>.
\end{equation} 
In view of the fact that the conformal factor is constant, we can change coordinates by 
$(\sqrt{\vp}\tilde r= r, \sqrt{\vp}\tilde t= t)$ and obtain
\begin{equation} 
g=  \(-A \>\d{\tilde t}^2 + \Frac[ \d{\tilde r}^2/ A] + \tilde r^2 \d\Om^2\)\>.
\end{equation} 
If we want a specific asymptotic behaviour, for example a metric which is asymptotically anti-de-Sitter, we can set $a=1$ in the function $A(r)$.

Since the solution of BD theory will be given in isotropic coordinates, let us first recast this solution in isotropic coordinates $(t, \rho, \te,\phi)$.
Any static, spherically symmetric metric 
\begin{equation} 
g =  -A(r) \d t^2 +C(r) \d r^2+ r^2 \d\Om^2 
\end{equation} 
can be recast in isotropic form
\begin{equation} 
g =  -A(\rho) \d t^2 +B(\rho) \( \d\rho^2+ \rho^2 \d\Om^2 \)
\end{equation} 
by a change of radial coordinate $\rho=\rho(r)$ (hence $\d\rho = \rho' \>\d r$).

\Note
One has simply
\begin{equation} 
g =  -A(\rho) \d t^2 +B(\rho)\> (\rho')^2 \d r^2+ B \rho^2 \d\Om^2
\end{equation} 
and comparing with the expression in pseudo-spherical coordinates one gets the conditions
\begin{equation} 
\begin{cases}
 B(\rho)\>  (\rho')^2 = C \\
B(\rho)\> \rho^2= r^2\\
\end{cases}
\quad\then
\(\Frac[\rho' / \rho]\)^2 = \Frac[C(r)/r^2]\>.
\end{equation} 
Hence one can integrate the last condition to get $\rho(r)$ and then set $A(\rho):= A(r(\rho))$.
\EndNote
For example, the Schwarzschild metric in pseudo-spherical coordinates $(t, r, \te,\phi)$ is 
\begin{equation} 
g= -\(\Frac[r-2m/r]\) \d t^2 + \Frac[ r / r-2m] \d r^2 + r^2 \d\Om^2
\end{equation} 
while in isotropic coordinates it reads as
\begin{equation} 
g= -\(\Frac[2\rho-m/2\rho +m ]\)^2 \d t^2 +  \(1+ \Frac[m/ 2\rho]\)^4 \(\d\rho^2 + \rho^2 \d\Om^2\)
\end{equation} 
where the integration constant has been fixed to have $\lim_{r\to +\infty} \Frac[\rho/r]=1$.
Let us remark that for $\rho\to +\infty$ we get Minkowski metric.

\subsection{Solution in BD theory}

We can find a static and isotropic solution of BD equations by an ansatz in isotropic coordinates $(t, \rho,\te, \phi)$, namely
\begin{equation} 
g= -A(\rho) \d t^2 + B(\rho)\( \d\rho^2 + \rho^2 \d\Om^2 \)
\quad\textrm{with}\quad
\vp = \vp(\rho)\>.
\end{equation} 
If we fix the potential to be zero and $\om\not=-\Frac[3/2]$, we can solve the BD equation \citep[see][]{Bhadra,Kozyrev} as

\begin{align}
A(\rho)&=\al_0 \(\Frac[ 2\rho-m / 2\rho+m ] \)^{\Frac[2/\la]}\>,\\
B(\rho)&= \be_0 \(\Frac[2\rho+m/2\rho]\)^4 \(\Frac[ 2\rho-m / 2\rho+m ] \)^{\Frac[2(\la-C-1)/\la]}
= \Frac[\be_0/\al_0^{\la-C-1}] \(\Frac[2\rho+m/2\rho]\)^4 A^{\la-C-1}\>,\\
\vp(\rho)&=\vp_0 \(\Frac[ 2\rho-m / 2\rho+m ] \)^{\Frac[C/\la]}\>,\label{BDSol}
\end{align}
where we have defined
\begin{equation}
\la^2=(C+1)^2-C+\Frac[\om/2] C^2=\(1+\Frac[\om/2] \)C^2+ C +1\>.
\end{equation}
That corresponds to what Weinberg does \citep{Weinberg}, though with no approximations.
It is a solution for any $(\al_0, \be_0, \vp_0, C)$ and $\la$ is computed with the identity above.
If we want to get the Schwarzschild metric at infinity, we need to set $(\al_0=\be_0=1)$.
Even considering $\al_0=\be_0=1$, we see that there is a 1-parameter family of static and spherically symmetric solutions, parameterized by $C$, unlike what happens in standard GR where the solution is unique.

We see that there are two families of solutions: 
one with $C=0$, and consequently $\la=\pm 1$, where the conformal factor $\vp$ is constant and indeed, when $\la=1$, reduces to the Schwarzschild metric.

\Note
\begin{equation} 
A(\rho)= \(\Frac[ 2\rho-\la m / 2\rho+\la m ] \)^{2}\>,
\quad\quad
B(\rho)= \(\Frac[ 2\rho+\la m/2\rho]\)^4 \>,
\quad\quad
\vp(\rho)=\vp_0\>.
\end{equation} 

\EndNote
The second family is for $C\not=0$ which has a non-constant conformal factor whenever $m\not=0$. Accordingly, we can say that Schwarzschild solution is always there, although as a somehow isolated solution, while BD theory allows a whole family of solutions with a non-constant conformal factor.

For $\om= -\Frac[3/2]$ and no potential, one has an arbitrary conformal factor and the Schwarzschild metrics only. 
If the potential is added as in view of the dynamical equivalence, then the conformal factor is frozen to be constant and the Schwarzschild-de Sitter solution is obtained.

In what follows we shall consider a sub-family of the general solution with $C=1/n$, and consequently $\la^2=[2(n^2+n+1)+\om ]/(2n^2)$
(which for $\om \to -\Frac[3/2]$ gives $\la\to \Frac[2n+2 / 2n ]$, while for $n\to \infty$ gives $\la=\pm 1$ which is the value associated to the Schwarzschild solution)
and we shall compute an observable, such as the precession rate of Mercury. We shall show that even in the limit $\om \to -\Frac[3/2]$
such an observable is discontinuous and it does not reproduce the result for the Schwarzschild-de Sitter solution obtained by setting $\om = -\Frac[3/2]$.
That eventually justifies the claim that the value  $\om = -\Frac[3/2]$ is degenerate and one cannot use a limit procedure to infer the result in the degenerate case,
and then neither in the dynamically equivalent Palatini $f(\cR)$-theory (which in fact will pass the Mercury test).

\section{Geodetics and exact relativistic Kepler laws}

We shall predict precession of Mercury by solving a more general problem, i.e.\ writing {\it exact Kepler laws} for an arbitrary static spherically symmetric metric
in isotropic coordinates restricted to the equatorial plane $\te=\Frac[\pi/2]$,
\begin{equation} 
g= -c^2 A(r) \d t^2 + B(r) \d r^2 + r^2 \d\phi^2 \>.
\end{equation} 
Precession comes from failure of first Kepler laws, since it measures by how much the orbits fail to be closed.
{\it Orbits} are $g$-timelike $g$-geodesics, in an allowed region where $r$ is bounded from below and above. 

We could use two Lagrangians. The first Lagrangian is for geodesics parameterised by proper time, i.e.\ 
\begin{equation} 
\tilde L = \Frac[m/2]\(c^2 A(r) \({\Frac[\d t/\d\tau]}\)^2 - B(r) \({\Frac[\d r/\d\tau]}\) ^2 - r^2  \({\Frac[\d\phi/\d\tau]}\)^2 \) \hbox{d}\tau\>.
\end{equation} 

\Note
This Lagrangian has 3  d.o.f. $(t, r, \phi)$ and 3 first integrals $(P, H, K)$, namely
\begin{equation} 
P =  A {\Frac[\d t/\d\tau]}\>,
\quad
K= r^2 {\Frac[\d\phi/\d\tau]}\>,
\quad
H=  c^2 A \({\Frac[\d t/\d\tau]}\)^2 - B \({\Frac[\d r/\d\tau]}\)^2 - r^2 \({\Frac[\d\phi/\d\tau]}\)^2 \>,
\end{equation} 
which we can solve as 
\begin{equation} 
{\Frac[\d t/\d\tau]} = \Frac[P /   A]\>,
\quad
 {\Frac[\d\phi/\d t]}= \Frac[ K A / Pr^2] \>,
\quad
 \({\Frac[\d r/\d t]}\)^2 = \Frac[\( c^2 P^2 - H\A \)r^2  - K^2\A/   P^2 AB r^2]  A^2\>,
\end{equation} 
\begin{equation} 
\({\Frac[\d r/\d\tau]}\)^2 =    \Frac[\( c^2 P^2 - HA \)r^2  - K^2A/  AB r^2]  \>.
\end{equation} 
\EndNote

The second Lagrangian is invariant with respect to reparameterisations
\begin{equation} 
L= mc \sqrt{c^2A(r)  - B(r)  \({\Frac[\d r/\d t]}\)^2 -r^2  \({\Frac[\d\phi/\d t]}\)^2 } \> \d t\>.
\end{equation} 

\Note
This Lagrangian has 2 d.o.f. $(r, \phi)$ and 2 first integrals $(E, J)$.
\hfill($[J]= L$, $[E]=TL^{-1}$)
\begin{equation} 
J=- \Frac[ r^2 {\Frac[\d\phi/\d t]} /\sqrt{c^2  A(r) - B(r)  \({\Frac[\d r/\d t]}\)^2 -r^2  \({\Frac[\d\phi/\d t]}\)^2 }]\>,
\quad
E=-  \Frac[  A / \sqrt{c^2 A(r) - B(r)  \({\Frac[\d r/\d t]}\)^2 -r^2  \({\Frac[\d\phi/\d t]}\)^2 }]\>.
\end{equation} 
We have
\begin{equation} 
{\Frac[\d\phi/\d t]} =  \Frac[ JA / Er^2 ]   \>,
\quad
  \({\Frac[\d r/\d t]}\)^2 = \Frac[  c^2   A /  B  ]  -  \Frac[  A^2 / E^2 B  ] -  \Frac[ J^2 A^2   / E^2 B r^2 ] 
   = \Frac[  c^2 E^2   r^2 - A (r^2 + J^2 ) /  E^2 AB  r^2] A^2\>.
\end{equation} 
Now we can map the values of the first integrals in the two frameworks
\begin{equation} 
 \Frac[K  / P] =  \Frac[J / E ]   \>,
 \quad
  E^2  =    \Frac[ P^2  / H]  \>.
 \end{equation} 
\EndNote

From the invariant Lagrangian we have
\begin{align}
\({\Frac[\d r/\d t]}\)^2 &=  \Frac[ A /   B  E^2 r^2 ]  \(c^2E^2 r^2   -A(r^2+J^2 ) \) =:\Phi(r; E, J)
\>,\\
\( {\Frac[\d\phi/\d r]}\)^2&= \Frac[ J^2 AB  / r^2 ]    \Frac[ 1  / c^2 E^2   r^2 -A (r^2 + J^2)  ] =:\Psi(r; E, J)\>,
\end{align} 
where the former equation gives the time evolution and the latter the orbit trajectory.

Having orbits is related to having an allowed region $[r_-, r_+]$ in $\Phi$ (as well as in $\Psi$). By integrating the latter in the allowed region $[r_-, r_+]$, one gets $2\phi= 2\pi+ \de$, where $\de$ is the precession per orbit. When the precession is zero, the orbit is closed, i.e.\ the classic first Kepler laws. Accordingly, the precession $\de$ is expected to be smaller and smaller getting away from the source, though not in the limit unless the solution is asymptotically flat (namely, $\La=0$). In the limit to standard GR, Mercury's precession should approach $\sim 43\,\mathrm{arcsec/century}$. For the Earth precession should be negligible.

The second Kepler law is related to conservation of angular momentum, which indeed is conserved exactly also in the relativistic regime.  Finally, we can get the period $T(r)$ from the integral in the allowed region. In the Newtonian approximation one has $T^2\propto (r_-+r_+)^3$. The function $T(r)$ contains the exact law, which in the limit must reproduce the Newtonian prediction.

Let us now make this explicit in some metrics which are relevant for the theories we are considering.

\subsection{Schwarzschild}
The Schwarzschild solution is defined as
\begin{equation} 
\A= 1- \Frac[b/r]= \Frac[r-b/r]<1
\quad\textrm{with}\quad
b:=\Frac[ 2G / c^2] m >0\>.
\end{equation} 
One has $[b]= L$ and $b$ is called the {\it Schwarzschild radius}.
\Note
Since the mass of the Sun is $M_\odot=1.9885\cdot 10^{30} \,\mathrm{kg}$, its Schwarzschild radius is $b= 2953.29\,\mathrm{m}$.
\EndNote

The function $\Phi$ is then given  by
\begin{equation} 
\Phi=    \Frac[  (r-b)^2 /  E^2 r^5 ]  \(( c^2 E^2  -1) r^3 + br^2 -J^2r  +bJ^2\)\>.
\end{equation}
This function to the power $-\frac{1}{2}$ will be integrated. In order to have an analytical expression of that integral it will be better to factorize the polynomial, i.e.~expressing it as a function of perihelion and aphelion $(r_\pm)$ instead of as a function of $(E, J)$.
If we want a bounded orbit, the function $\Phi$ must be negative at big $r$, thus $c^2 E^2  < 1$, i.e.\ $-1<cE<0$. The polynomial $p(r)=( c^2 E^2  -1) r^3 + br^2 -J^2r  +bJ^2 $ has a zero since it is of odd degree. We have $p(b)=( c^2 E^2  -1) b^3 + b^3 = c^2 E^2 b^3>0$ and  $-1<cE<0$, then $p(r)$ as a root $r_0>b$.

Hence, we can set initial conditions in the zero $r=r_0$, where $\dot r_0=0$, and set $\phi_0=0$, $\dot \phi_0= v_0/r_0$.
For these initial conditions, we have
\begin{equation} 
J=- \Frac[ r_0 v_0 /\sqrt{c^2  \A_0  -v_0^2   }]\>,
\quad
E=-  \Frac[  \A_0 / \sqrt{c^2 \A_0  -v_0^2   }]\>,
\end{equation} 
where we set $\A_0= \A(r_0)$.
For having a bounded orbit we must have $c^2E^2<1$, i.e.\ $v_0^2< \Frac[ b(r_0-b)/ r_0^2] c^2=\Frac[\xi_0b / (\xi_0+ b)^2] c^2= v_M^2$.

\Note

Let us also notice that $v_0^2< \Frac[ b(r_0-b)/ r_0^2] c^2=\Frac[\xi_0b / (\xi_0+ b)^2] c^2< c^2$.
Accordingly, if we start from zero initial speed $v_0=0$ and we increase it, at some point the orbit will become unbounded.
\EndNote

With these values of course we can express $\Phi(r, r_0, v_0)$ and factorise out of it $(r-r_0)$:
\begin{equation} 
\Phi(r)= (r-b)^2    \Frac[ (c^2 \A_0^2- c^2 \A_0+ v_0^2) r^3    + br^2(c^2 \A_0  -v_0^2) -r_0^2 v_0^2 r +b r_0^2 v_0^2   /  \A_0^2 r^5 ]  
\quad\then
\Phi(r_0)=0
\end{equation} 

\begin{equation} 
\Phi(r)= (r-b)^2 (r-r_0)   \Frac[ (c^2 b^2- c^2 r_0b + r_0^2v_0^2) r^2    + ( r_0-b)r_0^2 v_0^2 r -b r_0^3 v_0^2   /  (r_0-b)^2 r^5 ]
\>.  
\end{equation} 

If $\Phi$ has only one zero $r=r_0>b$, the allowed region is $[b, r_0]$ and the geodesic is falling towards the asymptotic goal at $r=b$, i.e.\ the horizon.
To have a bounded periodic orbit the allowed region must be $[b, r_0] \union [r_-, r_+]$.
Necessary condition for that to happen is 
\begin{equation} 
\De = ( r_0-b)^2r_0^4 v_0^4   -4 c^2 b^2(r_0-b) r_0^3 v_0^2+ 4b r_0^5 v_0^4   >0
\quad\then 
v^2_0 > \Frac[4(r_0-b) b^2 c^2/r_0 (b+r_0)^2]=: v_m^2\>.
\end{equation} 

\Note
Let us remark that $0< v_m^2<v_M^2 < c^2$.
In fact, $v_m^2$ is obviously positive. 
It is also
\begin{equation} 
v_m^2=  \Frac[4(r_0-b) b^2 c^2/r_0 (b+r_0)^2] = \Frac[4br_0 / (b+r_0)^2] v_M^2
=\Frac[b^2+ b\xi_0 / b^2+ b\xi_0+\({\Frac[\xi_0/2]}\)^2] v_M^2<  v_M^2\>.
\end{equation} 
Hence, starting from zero speed and increasing it, the test particle will fall in until it reaches a limit speed $v_m$ and it will stay bounded until a new limit $v_M$.
Accordingly, one has bounded orbits for $v_m^2<  v_0^2 <v_M^2$.
\EndNote

Then for $v_0= v_m$ we have a new zero of $\Phi$ appearing in $r_1=\Frac[2b r_0 / r_0-b]$,
which is  less that $r_0$, for any $r_0>2b$.

\Note
\begin{equation} 
\Frac[a(\xi_0+2b) / \xi_0] < \xi_0
\quad\then
-  \xi_0^2+ b\xi_0+2b^2 < 0\>.
\end{equation} 
One has that it vanishes for $\xi_0=b$ and it is negative for $\xi_0>b$ since it goes to $-\infty$ as $\xi\to \infty$. Accordingly, unless $b<r_0< 2b$, for $r_0>2b$ one has the first stable orbit (for $v_0=v_m$) in $\xi\in [\xi_1, \xi_0]$.
Then, increasing $v_0$ the perihelion grows to infinity. At some point il will reach and pass $\xi_0$, then it will grow to infinity which is reached at $v_0=v_M$.
\EndNote
When the perhelion becomes equal to the aphelion (circular orbit) and it keeps growing they get exchanged with each other. When we originally choose a zero $r_0$ of $\Phi$ we can always choose the biggest one. In this way we are interested to find the speed $v_c = v_0$
for which we have circular motion. That happens when $\Phi(r_0)=0$ twice, i.e.\ for $v_c^2= \Frac[c^2 b/2 r_0]$.

\Note
When $r_0>3b$, we have $v_m<v_c<v_M$.
Once again for $r_0<3b$ we are too near and Newtonian dynamics is not a good approximation.
\EndNote

Let us summarise. We choose initial position at the aphelion, for $r=r_+>3b$.\footnote{What happens for $r_+<3b$ is not our concern now.}
\begin{itemize}

\item[]{\it i)}
For $0<v_0< v_m$ (thus low angular momentum), the test particle falls in, end of the story.

\item[]{\it ii)}
For $v_0=v_m$, we have the first ``orbit'' with a perihelion of $r_-= \Frac[2b r_+ / r_+-b]<r_+$ (which is not an orbit yet, since $r_-$ is an asymptotic goal).

\item[]{\it iii)}
For $v_m^2<v_0^2< v_c^2=\Frac[c^2 b/2 r_+]$, we have elliptic orbits, with perihelion $\Frac[2b r_+ / r_+-b] <r_-(v_0)<r_+ $.

\item[]{\it iv)}
For $v_0^2= v_c^2$, we have circular orbits, $r_-=r_+$.

\item[]{\it v)}
For $v_c^2<v_0^2<v_M^2$, we have elliptic orbits, though the initial conditions are given in the perihelion, not in the aphelion. These orbits are recovered by giving initial conditions in the aphelion.

\item[]{\it vi)}
For $v_0^2=v_M^2$, one has parabolic orbits.

\item[]{\it vii)} For $v_0^2>v_M^2$, one has hyperbolic orbits.

\end{itemize}

\medskip

In what follows we are interested in $iii)$ and $iv)$ only since they capture all bounded orbits $r\in [r_-, r_+]$, 
parameterised by $r_+$ and $r_-$.
By the way, if we fix $r_-$ and $r_+$ the initial velocity we need for that is $v_0^2= \Frac[ c^2 b r_-^2  (r_+-b)/ r_+^2 (r_++r_- )( r_--b)]$. The first integrals for the bounded orbits are
\begin{align} 
cE&= -\sqrt{\Frac[(r_+-b)(r_--b)(r_++r_-) / r_-^2 r_+ +r_- r_+^2-b r_+ r_- -br_+^2  - br_-^2 ]}\>,\\
J&=  -\sqrt{\Frac[b r_+^2r_-^2 / r_-^2 r_+ +r_- r_+^2-b r_+ r_- -br_+^2  - br_-^2 ]}\>.
\end{align} 

Accordingly, we can express the function $\Phi$ in terms of $r_\pm$ as
\begin{align} 
\Phi(r, r_\pm)&= -bc^2 (r-b)^2\Frac[(r_+-r)(r- r_-)\( (br_-      -r_-r_+    +br_+)r    +br_+r_-\) / r^5(r_--b)(r_+ +r_-)(r_+-b)]\>,\\
\Psi(r, r_\pm)&= - \Frac[ r_+^2 r_-^2/ r (r_+-r)(r-r_-)\( (br_-      -r_-r_+    +br_+)r    +br_+r_-\)]\>.
\end{align} 

Alright then, we have expressed everything in terms of $r_\pm$.

\subsection{Exact Kepler laws in Schwarzschild}

For the Earth, we have $r_-=147095000000\,\mathrm{m}$ and $r_+=152100000000\,\mathrm{m}$. Then its period is
\begin{equation} 
T_{\earth}= 2 \int_{r_-}^{r_+} \Frac[ \d r/ \sqrt{\Phi(r; r_\pm)}]= 3.15578845707534\cdot 10^7 \,\mathrm{s}  = 1.00069\,\mathrm{yr}
\>.
\end{equation} 

\Note
This is {\it computed}, theoretical quantities, not measured. We are not meaning we can observe the Earth period up to $10^{-7}s$.
We mean that we can predict its value with arbitrary prediction so that we can compare observed value with it.
\EndNote
The Earth precession per orbit is
\begin{equation} 
\de_{\earth} = 2\(\int_{r_-}^{r_+} \sqrt{\Psi(r; r_\pm)} \> \d r-\pi\) = 1.8611182\cdot 10^{-7}\,\mathrm{rad}\>,
\end{equation}  
and consequently, the precession per century (i.e.\ the cumulative precession for 100 Earth's periods $T_{\earth}$) in $\mathrm{arcsec}= \mathrm{deg}/3600$ is $\de_{\earth}=3.839\,\mathrm{arcsec}$.

On the other hand, for Mercury we have $r_-=46001200000\,\mathrm{m}$ and $r_+=69816900000\,\mathrm{m}$. Then its period is 
\begin{equation} 
T_{\mercury}= 2 \int_{r_-}^{r_+} \Frac[ \d r/ \sqrt{\Phi(r; r_\pm)}]= 0.760048069820773\cdot 10^7 \,\mathrm{s} =0.241010\,\mathrm{yr}\>.
\end{equation} 
The Mercury's precession per orbit is
\begin{equation} 
\de_{\mercury} = 2\(\int_{r_-}^{r_+} \sqrt{\Psi(r; r_\pm)} \> \d r-\pi\) =5.0187261\cdot 10^{-7}\,\mathrm{rad}\>,
\end{equation}  
implying a precession per century (i.e.\ the cumulative precession for 100 Earth's periods $T_{\earth}$, which corresponds to about $415.21$
Mercury's periods) of $\de_{\mercury} = 42.98\, \mathrm{arcsec}$.

Let us stress that comparing with Earth's period to define {\it century} allows us to avoid any direct reference to an external clocks.
In some sense we are using relational time within the system itself.

To find third law we can consider $\Phi$ and $\Psi$, make the substitutions $r\to \rho/x$, $r_\pm\to (1\pm e)/x$ and expand at $x=0^+$ (i.e.\ far away from the central mass). The first term in the series reproduces the corresponding Kepler function. The second term in the series gives corrections. This is a very convenient technique to define what is the Newtonian limit, since it is done {\it before} integration. Of course, it requires one does Kepler case first.

 \subsection{Post Newtonian approximation}

One usually does is expanding the metric coefficients in isotropic coordinates in series of $MG/\rho$ and assume it is not very different from Minkowski, i.e.\ the weak field approximation, namely
\begin{equation} 
g= -\(1-2\al \Frac[MG/\rho] + 2 \be \Frac[M^2G^2/\rho^2]  \) \d t^2+
\(1+2\ga \Frac[MG/\rho] \) (\d\rho^2 + \rho^2 \d\Om^2)\>.
\end{equation}  
This approximation is good enough for Mercury, it would fail too close to a black hole.
In this approximation the Schwarzschild solution is recovered for $\al=\be=\ga=1$, while the BD solution is
obtained for 
\begin{align} 
\al&=\be=1,\\
\ga&=\Frac[\om+1/\om+2] \>.
\end{align}  

It is clear that in the very same moment one expands in series, the ability to spot isolated singular solutions is lost.
Of course, for  $\om\to \infty$, one gets $\ga=1$ for the Schwarzschild solution, but for $\om=-3/2$ one gets
$\ga= -1$.
As a matter of fact, one is assuming {\it a priori} to be on the main regular sequence of solution,
which is incorrect (or a partial viewpoint) as we shall see. For this reason it is much better to stick to exact results rather than starting expanding in series. As an extra bonus, by doing it exactly, we also can test how close we need to be to see the strong field effects which, in principle, is a solid prediction of the theory.

\subsection{Schwarzschild-de Sitter}

Let us discuss the Schwarzschild-de Sitter spacetime with $A= 1 -b/r - \la r^2$.
That is a solution of Einstein equation in vacuum with cosmological constant $\La= 3\la$. We need $A>0$ at least in a region $r\in [r_1, r_2]$. Since we have
\begin{equation} 
A= 1 -\Frac[b/r] - \la r^2= \Frac[ -\la r^3+ r-b / r]\>,
\end{equation} 
if $\la>0$, we have $\lim_{r\to +\infty} A=-\infty$ and, if $b>0$, we have $\lim_{r\to 0} A=-\infty$.
Thus we would like a region in the middle where $A>0$, which is easy to find since 
\begin{equation} 
A'=  \Frac[b/r^2] - 2\la r  = \Frac[-2\la r^3+ b / r^2]=0
\quad\iff\quad
r_\ast={}^3\<\<\sqrt{\Frac[ b / 2\la]}
\end{equation} 
and in $r=r_\ast$ we have
\begin{equation} 
A= \Frac[ -\la {\Frac[ b / 2\la]}+ {}^3\<\<\sqrt{\Frac[ b / 2\la]}-b /{}^3\<\<\sqrt{\Frac[ b / 2\la]}]
=\( 1 - \Frac[3/2]\> {}^3\<\<\sqrt{ 2b^2\la}  \) >0
\quad\then\quad
0<\la< \Frac[4/27b^2] \>.
\end{equation} 

\Note
If $\la$ is too big there is no such a region where $A>0$. That means that as $\la$ grows, sooner or later the cosmological horizon will touch the 
Schwarzschild one.
\EndNote
Thus the cosmological constant has to be small enough.
In this case the function $A$ has at least two zeros, hence three (since it is odd degree).
\begin{equation} 
\begin{aligned}
A=& \Frac[ -\la(r-r_0)(r-r_1)(r- r_2)  / r]=\\
=&\Frac[ -\la r^3 +\la(r_0+ r_1+ r_2)r^2 -\la(r_0r_1+ r_0r_2+ r_1r_2)r+\la r_0r_1r_2  / r]\>.
\end{aligned}
\end{equation} 
Hence, we have $r_0=-( r_1+ r_2)$ and 
\begin{equation} 
\begin{aligned}
A=&\Frac[ -\la r^3  +\la(r_1^2 +r_1r_2 +r_2^2)r-\la (r_1^2r_2+r_1r_2^2)  / r]
\quad\then\quad\\
\quad&\then\quad
\begin{cases}
 \la= \Frac[ 1/ r_1^2 +r_1r_2 +r_2^2]\\
 b=  \Frac[ r_1^2r_2+r_1r_2^2/ r_1^2 +r_1r_2 +r_2^2]\\
\end{cases}\>,
\end{aligned}
\end{equation} 
which is solved for $(r_1(\la, b), r_2(\la, b))$.

In what follows, we shall fix the (positive, small) {\it Schwarzschild radius} $r_1$ and the 
(positive, large) {\it cosmological radius} $r_2$, thus setting $r_0=-r_1-r_2$. Thus, we can write
\begin{equation} 
A= \Frac[ -\la(r-r_0)(r-r_1)(r- r_2)  / r]
= \Frac[ \(r+(r_1+r_2)\)(r-r_1)(r_2-r)  / r (r_1^2 +r_1r_2 +r_2^2)]\>,
\end{equation} 
which indeed is positive in the allowed region $r\in[r_1, r_2]$,
and we compute the corresponding $(a, \la)$ out of $(r_1, r_2)$.
The function $\Phi$ is then given  by
\begin{align}
\Phi&=  A^2   \( c^2     -  \Frac[r^2+J^2  /  E^2 r^2 ] A \)\\
&=   \Frac[ c^2E^2 (r_1^2 +r_1r_2 +r_2^2)r^3 -(r^2+J^2)(r+r_1+r_2)(r-r_1)(r_2-r) / (r_1^2 +r_1r_2 +r_2^2)^3E^2 r^5] 
\cdot\\&\cdot 
(r+r_1+r_2)^2(r-r_1)^2(r_2-r)^2\>,
\end{align}
while the function $\Psi$ is given by
\begin{equation} 
\Psi=   \Frac[J^2/r] \cdot  \Frac[   r_1^2 + r_1r_2 +r_2^2 /  c^2E^2 (r_1^2 +r_1r_2 +r_2^2)r^3 -(r^2+J^2)(r+r_1+r_2)(r-r_1)(r_2-r) ] \>.
\end{equation} 

Thus, within the region $[r_1, r_2]$ the behaviour of $\Phi$ (and $\Psi$), in particular the zeros and the allowed regions, are ruled by the $5$th-degree polynomial 
\begin{equation} 
p(r)=c^2E^2 (r_1^2 +r_1r_2 +r_2^2)r^3 -(r^2+J^2)(r+r_1+r_2)(r-r_1)(r_2-r)\>.
\end{equation} 
If we want a bounded orbit, $p(r)$ must have 4 zeros,
$\{r_m, r_+, r_-, r_M\}$, in the region $[r_1, r_2]$ so one negative zero $r=-(r_m+ r_++r_- + r_M)$.
Hence it is
\begin{equation} 
p(r)=(r+r_m+ r_++r_- + r_M)(r-r_m)(r-r_-)(r-r_+)(r-r_M)\>.
\end{equation} 
If one knows the mass of the star and cosmological constant, then $r_1$ and $r_2$ can be computed.
Then one gives a planet with its $r_\pm$ and can compute out of $(r_1, r_2, r_\pm)$ the value of $(E, J, r_m, r_M)$, i.e.\ the initial conditions.
In other words, while $(r_1, r_2)$ are a convenient way of parameterizing the parameters of the system, namely $(m, \la)$, 
$(r_+, r_-)$ are a convenient way of parameterizing initial conditions of a specific timelike geodesic.

\Note
Although it is good to discuss it once from scratch, one can also cut the discussion short by saying that we want to have an orbit, i.e.\ time-like geodesic that is bounded from above and below. That means we need an allowed region $[r_-, r_+]$ for $\Phi$ and $\Psi$, so that both $r_\pm$ are simple zeros. Knowing that, at infinity, $\Phi$ is definite negative and it is positive around $r_1<r_m<r_-\le r_+< r_M< r_2$, we directly get
\begin{equation} 
\begin{aligned}
\Phi =&   \Frac[ (r+r_m+ r_++r_- + r_M)(r-r_m)(r-r_-)(r-r_+)(r-r_M)/ (r_1^2 +r_1r_2 +r_2^2)^3E^2 r^5] 
\cdot\\
\cdot& (r+r_1+r_2)^2(r-r_1)^2(r_2-r)^2\>.
\end{aligned}
\end{equation} 
Note that we also have two allowed regions $[r_1, r_m]$ and $[r_M, r_2]$ corresponding to the test particle falling in and escaping to infinity, respectively. Anyway, here we are interested in solutions in $[r_-, r_+]$.

Let us finally remark that the factorised form of $\Phi$ is very convenient for analytical computation. Most of the the possibility of treating the problem analytically relies on this factorisation, i.e.\ in using $(r_1, r_2, r_\pm)$ instead $(\la, m, E, J)$ and express the rest in terms of them.
\EndNote

Now that $\Phi(r)$ and $\Psi(r)$ are fixed, one can compute the planet period and the precession per orbit as we did for Schwazschild. That can be done for Earth and for Mercury so to have the ratio between periods. Then we can compute the precession of Mercury in a century, obtaining
\begin{equation} 
\begin{aligned}
\La=&\La_+=  1.2729\cdot 10^{-52}  \,\mathrm{m}^{-2}
\\
\de_{\mercury} =& 42.9818839109594321936164983375 \,\mathrm{arcsec\,century^{-1}}
\end{aligned}
\end{equation} 
to be compared with the result in GR
\begin{equation} 
\begin{aligned}
\La=& 0
\\
\de_{\mercury} =& 42.9818839109594 2778584505400996\,\mathrm{arcsec\,century^{-1}}\>.
\end{aligned}
\end{equation} 

Accordingly, we have a relative error of $\De=10^{{-16}}$ if we neglect the cosmological constant.

\Note
We can repeat the computation for different values of the cosmological constant to check how it grows when we switch it on.
In fact we have
\begin{equation} 
\begin{aligned}
\La= 0 				&\quad\quad \de_{\mercury} = 42.9818839109594 2778584505400996   \quad\quad \De\sim 10^{-16} \\
\La= \La_+			&\quad\quad \de_{\mercury} = 42.9818839109594\,\vert\,321936164983375  \\
\La= 10\cdot \La_+		&\quad\quad \de_{\mercury} = 42.9818839109594\,\vert\,718635618144828 \quad\quad \De\sim 10^{-15}  \\
\La= 10^{2}\cdot \La_+	&\quad\quad \de_{\mercury} = 42.981883910959\,\vert\,8685630175452256 \quad\quad \De\sim  10^{-14} \\
\La= 10^{4}\cdot \La_+	&\quad\quad \de_{\mercury} =  42.98188391\,\vert\,10035055030691353771\quad\quad \De\sim 10^{-12}  \\
\La= 10^{8}\cdot \La_+	&\quad\quad \de_{\mercury} =  42.98188\,\vert\,43517366000242900907766 \quad\quad \De\sim 10^{-8}  \\
\La= 10^{11}\cdot \La_+	&\quad\quad \de_{\mercury} =  42.98\,\vert\,23246881316750018084036904 \quad\quad \De\sim 10^{-5}  \\
\La= 10^{12}\cdot \La_+	&\quad\quad \de_{\mercury} =  42.98\,\vert\,62916826826901188288108750 \quad\quad \De\sim 10^{-4}  \\
\La= 10^{13}\cdot \La_+	&\quad\quad \de_{\mercury} =  43.0259616282710684507382369547 \phantom{\,\vert\,}\quad\quad \De\sim 10^{-3}  \\
\La= 10^{14}\cdot \La_+	&\quad\quad \de_{\mercury} =  43.4226610919775681642491924622 \phantom{\,\vert\,}\quad\quad \De\sim 10^{-2}  \\
\La= 10^{15}\cdot \La_+	&\quad\quad \de_{\mercury} =  47.3896565113144280853078911238 \phantom{\,\vert\,}\quad\quad \De\sim 10^{-1}  \\
\La= 10^{16}\cdot \La_+	&\quad\quad \de_{\mercury} =  87.0596889320926516513462746807 \phantom{\,\vert\,}\quad\quad \De\sim 10^{0}  \\
\La= 10^{18}\cdot \La_+	&\quad\quad \de_{\mercury} = 4451.63182768107067978911658191 \phantom{\,\vert\,}\quad\quad \De\sim 10^{2}  \\
\end{aligned}
\end{equation} 
which shows that the effect of the cosmological constant grows approximately linearly in the $(\log,\log)$-graph
and that with a cosmological constant $\La= \La_+$ we are well within the limit in which we cannot observe it in Mercury perihelion. We put 
a bar to highlight on its left hand side the digits which agrees with the value computed with no cosmological constant. 
That bar also highlights on the right hand side the digits which are affected by the cosmological constant.
Let us notice that we are not simply saying that as long as we consider Mercury's precession we can neglect the cosmological constant.
We are in fact computing the relative error one does by neglecting it.
\EndNote

\subsection{Solution in $f(\cR)$}

We have considered the Schwarzschild-de Sitter metric and computed precession of Mercury in them.
We found that if the cosmological constant is small enough the theoretical precession is compatible with the observed one.

In view of universality theorem we know that (vacuum) Palatini $f(\cR)$-theories in fact are equivalent to Einstein with a cosmological constant. However, extra care is needed in this case. Universality theorem claims that $\tilde g$ is a solution of Einstein equations with a cosmological constant the value of which is dictated by the function $f(\cR)$ via the master equation. 
Now from the viewpoint of Ehlers Pirani and Schild (EPS) framework (see \cite{EPS}, \cite{EPSNostro}), one should expect $\tilde g$ to govern geodesic motions, while $g$ is related to distances, while in Schwarzschild-de Sitter model (as above) one has a single metric, namely $g$ above, which dictates both geodesic equations and distances. Although in principle one should discuss whether this aspect plays a relevant role when applying the discussion above as it does in general, we have to remark that Solar System is modelled by a {\it vacuum} solution of Palatini $f(\cR)$-theory, in which hence the conformal factor is constant. Accordingly,  $\{g\}= \{\tilde g\}$, i.e.\ the two metrics $g$ and $\tilde g$ actually define the same geodesics trajectories, the same timelike directions, and one has 
$\tilde R_{\mu\nu}=R_{\mu\nu}$, i.e.\
\begin{equation} 
\tilde R \tilde g_{\mu\nu}
=\tilde g^{\rho\si} \tilde R_{\rho\si}  \tilde g_{\mu\nu} 
= g^{\rho\si} \tilde R_{\rho\si}   g_{\mu\nu} = \begin{cases}
\cR g_{\mu\nu} \\
g^{\rho\si}  R_{\rho\si}   g_{\mu\nu} = R g_{\mu\nu}  \>.\\
\end{cases}  
\end{equation} 
Accordingly, also $g$ obeys the same field equations as $\tilde g$.
We can use $g$ only in vacuum, and apply the result above.
 
If the function $f(\cR)$ determines a small enough cosmological constant (as it happens with the function (\ref{fR}) and parameters (\ref{PintoPar})) then it actually predicts precession of Mercury compatible with the observed one. Let s remark that Mercury test is passed despite the value $\al\simeq 0.095$. That directly shows that $\al \simeq 1$ is not required to pass this test. In view of dynamical equivalence between Palatini $f(\cR)$ theory and BD (with a potential and $\om=-3/2$) this shows also that such a BD theory passes 
the test as well.

Also in this case, one should pay attention to the fact that in Palatini $f(\cR)$-theories geodesics and distances are related to two different metrics while in BD
both are related to the same metric $g$. However, in vacuum, this is not an issue since $g$-timelike $\tilde g$-geodesics are also (and the only) $g$-timelike $g$-geodesics. However, in non-vacuum solutions (as galactic dynamics or cosmology) the models would be actually different.

Since classical tests have been performed in BD theories (with no potential and generic $\om$, hence different from $\om=-3/2$, which is degenerate) and they show that $\om$ must be $\om>10^4$, this was used to try, erroneously as we discussed above, to rule out all Palatini $f(\cR)$-theories at once. Besides, we saw directly that this argument is spurious, we shall review the test in BD model on an exact formulation. We have two reasons to do it: first, since we will not use PN approximation for it we can apply it close to the horizon, in the strong field regime. Secondly, we show that the value $\om=-\Frac[3/2]$ is degenerate also with respect to the test and the role of the potential cannot be neglected.

\section{Mercury test in BD theory}

Let us here consider a BD theory with no potential (and $\om\not=-3/2$, as this is used to determine the solution as long as the scalar field is concerned). Test particles (and the planets Earth and  Mercury) are assumed to go along geodesics of $g$. On the equatorial plane, in isotropic coordinates, one has the Lagrangian
\begin{equation} 
L_{geo} =\mu c \sqrt{c^2 A(\rho)- B(\rho) \( \(\Frac[\d\rho/\d t]\)^2  + \rho^2  \(\Frac[\d\phi/\d t]\)^2\) } \> \d t\>,
\end{equation} 
again with total energy and angular momentum as first integrals. In isotropic coordinates, the coefficients are a bit different to what described so far, so we need to repeat the discussion from scratch.

First integrals are
\begin{align} 
J&=- \Frac[ B(\rho) \rho^2 {\Frac[\d\phi/\d t]} /\sqrt{c^2  A(\rho) - B(\rho)  \({\Frac[\d\rho/\d t]}\)^2 -\rho^2 B(\rho) \({\Frac[\d\phi/\d t]}\)^2 }]\>,\\
E&= -  \Frac[ A(\rho) / \sqrt{c^2 A(\rho) - B(\rho)  \({\Frac[\d\rho/\d t]}\)^2 -\rho^2 B(\rho) \({\Frac[\d\phi/\d t]}\)^2 }]\>,
\end{align} 
which can be inverted for velocities as
\begin{align}
\({\Frac[\d\rho/\d t]}\)^2 &=  \Frac[ A /   B^2  ]  \( c^2 B     - A \Frac[ B \rho^2+J^2  / E^2 \rho^2 ] \)=\Phi(\rho; E, J; m, C, \om)\>,\\
\({\Frac[\d\phi/\d t]}\)^2 &=  \Frac[ J^2 A^2/ B^2\rho^4 E^2 ]   
\quad\then
\({\Frac[\d\phi/\d r]}\)^2 =  \Frac[J^2 A/ { \rho^4 E^2   \( c^2 B     - A \Frac[ B \rho^2+J^2  / E^2 \rho^2 ] \)} ] =\Psi(\rho; E, J)\>.
\end{align}
Considering $\Phi$, if $\rho\to +\infty$, given that $A\to 1$ and $B\to 1$, we have 
\begin{equation} 
\Phi\to   \Frac[ c^2 E^2 -1 / E^2  ] <0 
\quad\then
c^2 E^2 < 1
\quad\then  
-1<cE\le 0
\end{equation} 
in order to have bounded orbits.

Now that we know how it works, we can either study the function $\Phi$ for all parameters and then select parameters which describe bounded orbits, or simply require two solutions $\Phi(\rho_\pm)=0$ that bound an allowed region $[\rho_-, \rho_+]$. Indeed, if we consider the equations $\Phi(\rho_\pm; E, J)=0$, we can solve for $\(E(\rho_\pm), J(\rho_\pm)\)$ and then replace them back into the Weierstrass functions to obtain
\begin{align} 
\(\Frac[\d\rho / \d t] \)^2&= \Phi(\rho; \rho_\pm)\>,\\
\(\Frac[ \d\phi / \d\rho] \)^2&= \Psi(\rho; \rho_\pm)\>.
\end{align} 

The second issue to discuss is how to use orbital parameters. Since we are in isotropic coordinates, the coordinate $\rho$ is not endowed with a direct meaning of a distance and it is different from the coordinate $r$ in pseudo-spherical coordinates \cite[see][]{Distances}. In view of the form of the metric in spherical coordinates, spacetime is foliated into Euclidean spheres (we mean {\it metric spheres}, not only topological spheres) arameterised 
by $(t, r)$, with the coordinate $r$ being the radius each sphere would have if it were embedded into a Euclidean space. That in fact corresponds to the observation protocol for measuring astronomical distances, which in fact uses Newtonian approximation and classical Kepler laws. Accordingly, the observed orbital parameters have to be related to $r$, not to $\rho$.
However, we know that 
\begin{equation} 
B(\rho) \rho^2= r^2
\label{IsotropicAphelion}
\end{equation} 
and we can determine the values of $\rho_\pm$ which correspond to the observed $r_\pm$ for Mercury or the Earth.

In order to consider the solution (\ref{BDSol}) in BD theory, we know that standard GR corresponds to a big value of $\om$, $C=0$ and $\la=1$.
In view of the constraint among $(\om, \la, C)$ we can consider a sub-family of solutions given by setting $\la=1$, $C=-\Frac[1/n]$
and computing $\om= 2(n-1)$ and hence table the exact precession of Mercury for each value of $n$.

This gives us a (partial) insight on how precession of Mercury depends on $(\om, C)$ in BD theory. We expect to obtain a good agreement with observed values for 
$n\to \infty$ and observe rather incompatible values for small values of $n$.
We can also consider the limit $n\to \Frac[1/4]$, which corresponds to $\om\to -\Frac[3/2]$, i.e.\ the degenerate parameter although with no potential.

The whole computation of the precession for a specific value of $n$ involves exact calculations up to a finite number of numerical integrations of converging improper integrals.
This is not much different from what one does when studying the graph of a transcendent function, in which a finite number of evaluations of the function 
(e.g.\ to determine zeros or critical points) are eventually performed numerically. Accordingly, we can say our computation is an {\it analytical exact result} or, if you prefer, call it {\it semi-analytical}.

For each value of $n$ we computed the orbital period of the Earth, of Mercury, the precession per orbit of Mercury and finally obtain the precession $P(n)$ of Mercury
during 100 orbits of the Earth. 

\begin{equation} 
\begin{aligned}
n 		\quad\quad
		& P(n)\> (arcsec/century)\\
\Frac[1/4]	\quad\quad
		&\ \kern-15pt  -71.64				\\
1		\quad\quad
		&	 14.33				\\
4		\quad\quad
		&	 35.82				\\
10		\quad\quad
		&	 40.12				\\
50		\quad\quad
		&	 42.41				\\
100		\quad\quad
		&	 42.70				\\
500		\quad\quad
		&	 42.92				\\
1000		\quad\quad
		&	  42.95\>.				\\
\end{aligned}
\label{PrecessionN}
\end{equation} 

These values are, theoretical predictions in different solutions of BD theories. They can be computed at arbitrary precision. Here we checked that the shown digits are
not affected when the overall precision required is increased.

We see that in fact one can distinguish among different values of $n$ (and $\om$) by means of $P$ which is definitely observable.
As the observed value of about $42.98\,\mathrm{arcsec\,century^{-1}}$, observations are able to exclude small values of $\om$, including $\om=-\Frac[3/2]$, as expected.

For the degenerate value $\om=-\Frac[3/2]$ one has two possible solutions, one for $C=\Frac[1/4]$ which sits in the sequence we analyzed and one for $C=0$
which does not and it is corresponds to ordinary Schwarzschild.
In fact, for the solution with $C=0$ the value of $\om$ is undetermined and the corresponding solution is somehow isolated in the solution space.
It passes the tests obviously, since it is the same solution of standard GR. 

Accordingly, it is not completely correct to say that BD theories with  small value of $\om$, even with no potential, are ruled out by observations. 
The solutions with $C\not=0$ are, while of course $C=0$ is not.

\subsection{Constraints in parameter space}

We consider BD solutions for  $\om$ and $C$.
We sample parameter spaces computing precession of Mercury (in $\mathrm{arcsec\,century^{-1}}$) and computing $\De P(\om, C)$ the relative error ($\cdot 100$) with respect to the observed value of about $42.98045118132$.

\begin{figure}[htbp] 
   \centering
   \includegraphics[height=12cm]{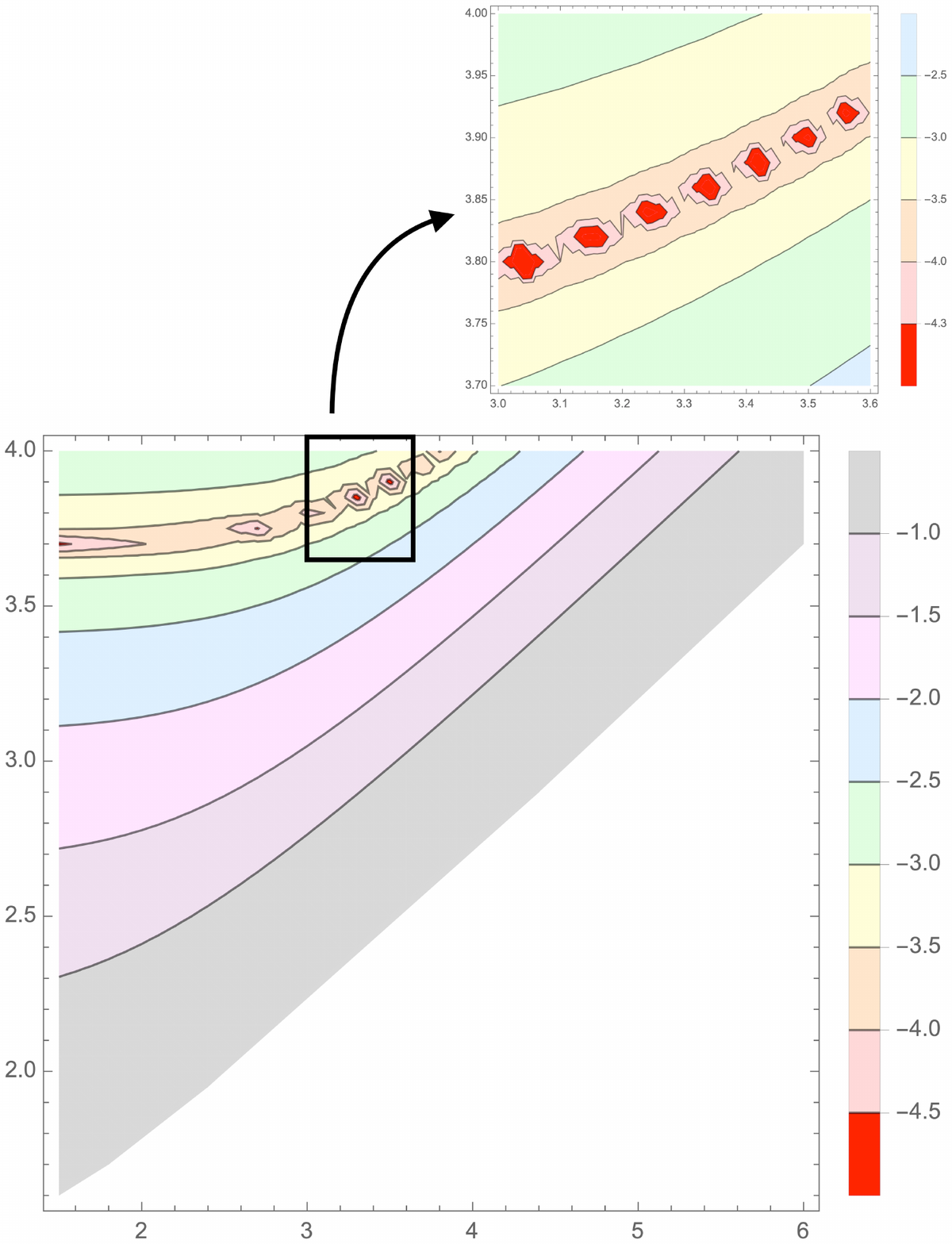}
   \caption{\it Relative error between computed precession and observed value ($\log$-scale) as a function of $\log(\om)$ on $x$-axis 
   and $-\log(-C)$ on $y$-axis.
   Standard GR corresponds to $\om\to +\infty$ and $C\to 0$, hence it corresponds to the direction $(+\infty, + \infty)$ in the plane. }\label{fig:contourplot}
\end{figure}
The values of $\om$ and $C$ which have interestingly low errors are too spread in the $(\om, C)$ plane.
It is expected low errors in the limit $\om\to +\infty$ and $C\to 0$ that corresponds to standard GR in the non-degenerate sequence.
Thus we plot $\De P(\om, C)$ on the axes $x=\log(\om)$, $y=-\log(C)$, so that standard GR corresponds to the limit
$x\to +\infty$ and $y\to +\infty$.
Indeed, we see in Fig.~\ref{fig:contourplot} that in the region $x\to +\infty$ and $y\to +\infty$ one has the smallest errors, as expected.


\subsection{Conformal factor and Weyl transformations}
We can still explore one possibility. When we used aphelion and perihelion in isotropic coordinates we used the metric $g$
as EPS framework dictates. However, we have two metrics and we should check the difference with what is predicted if we used 
$\tilde g$ instead.
This is relatively easy, all we have to do is determining $\rho_\pm$ by using the equation
\begin{equation} 
\vp(\rho) B(\rho) \rho^2 = r_\pm^2
\end{equation} 
instead of (\ref{IsotropicAphelion}).
In the case $n=1000$, we obtain the precession prediction to be
\begin{equation} 
\tilde P(1000) =  	42.95\,\mathrm{arcsec\,century^{-1}}
\end{equation} 
to be compared with the value computed in (\ref{PrecessionN}), namely $P(1000)= 42.95\,\mathrm{arcsec\,century^{-1}}$.
Thus we see a tiny difference, a difference yet, as tiny as expected since the conformal factor is very close to $1$ at the orbit of Mercury or the Earth.

Still, the difference is there and is expected to become bigger in stronger regimes which, by the way, says that the difference between using $g$ or $\tilde g$ to describe distances is {\it in principle} observable.

\section{Conclusion and Perspectives}

We considered the standard test of Mercury in different contexts.
Our treatment is analytical and we do not resort to weak field approximations so that our framework is still valid for satellites orbiting a black hole at few Schwarzschild radiuses.

It has been argued that since BD theory is ruled out by observations (for small values of $\om$) and they are dynamically equivalent to Palatini $f(\cR)$-theories,
then these are ruled out as well.

We showed that this is not the case for a number of reasons. First of all, dynamical equivalence is with a degenerate value of the parameter $\om=-\Frac[3/2]$, moreover with a potential which has not been considered in the original BD test.
Moreover, in BD theory one has no Birkhoff theorem so one has a three parameters' family of static spherically symmetric solutions.

We considered standard GR, standard GR with a cosmological constant, Palatini $f(\cR)$-theories and BD theories.
These produce Schwarzschild solution with or without a cosmological constant $\La$ as well as a more general family (\ref{BDSol}) of solutions of BD theory.

We showed that Schwarzschild, as well as Schwarzschild-de Sitter solutions pass the test provided that the cosmological constant $\La$ is small enough.
For Palatini $f(\cR)$-theories this imposes constraints on the function $f(\cR)$ which for example are met in some of the models based on (\ref{fR}),
among which one has the best fit values (\ref{PintoPar}) found in \cite{Pinto1}.
Among solutions of BD theory we showed that, besides Schwarzschild solution which is also a solution of BD theory for any value of $\om$ and which passes the test, 
also the other solutions of BD theory passes the tests provided that $\om$ is big enough.

This shows directly how Mercury test does not technically rule out BD for small $\om$, it rules out some solutions of it. 
Of course, the Schwarzschild solution cannot be ruled out, rather one does not {\it need} BD theory to have a Schwarzschild solution.

It also shows directly that dynamical equivalence is irrelevant for ruling out Palatini $f(\cR)$-theories which in fact are ruled out only if they produced too big values of the effective cosmological constant.
In particular, in the family (\ref{fR}) considered in \cite{Pinto1} where the best fit value (\ref{PintoPar}) of $(\al, \be, \ga)$ were found to model $Ia$ supernovae,
the parameter fit were found to be strongly degenerate. In particular, the value of $\be$ was found to be poorly localised by $SNIa$ data (as it can be expecetd since the $\be$ parameter has very tiny effect in a universe where supernovae can occur) and also for any given value of $\al$ one could find a best value for $\ga$ which produces a good agreement with observations.

During the peer review of \cite{Pinto1} it has been argued that the best fit value of $\al \simeq 0.1$ would fail to model Solar System.
Here we showed that this is not the case. The ruling out of theories has to be carried over at the level of observables not of actions.
Despite $\al \simeq 0.1$ produces a vacuum action which does not approximate the standard GR vacuum action, the constant factor has no effect on observables
in vacuum, hence in the Solar System tests. This is also implied by universality theorem for Palatini $f(\cR)$-theories.

In fact, if $SNIa$ fix a best fit value for $\ga$, then Mercury test produces constraints for the ratio $\Frac[\ga/\al]$.
The two set of data in fact remove the degeneracy for $\al$ and $\ga$, leaving the one connected to $\be$.

As for future perspectives, we need to extend the analysis to the other classical Solar System test (light deflection and radar delays)
to check whether these add constraints. We need to extend our exact approach to these cases so that conclusions are robust against weak field approximations and 
stay valid in a strong field regime.

Further constraints may arise from tight binary systems in which the weak field approximations can be at stake, as well one can look for consequences in collapse events relevant to gravitational waves. Being our method viable for different theories this would open a way to use gravitational wave phenomenology to be used for reliably distinguish different gravitational theories, especially when much more precise data will be available with new experimental surveys; see e.g.\ \cite{Euclid}, \cite{HD}.

Under this viewpoint of distinguishing different theories in terms of observables only, this paper is in a series with the aim of discussing validity of a specific family of theories (namely, in this case (\ref{fR})). 
Before even discussing the validity of a model, one needs to fix parameters. 
This is not different from what it is routinely done in QFT, when a general theory of electromagnetic field and its interaction with charged fields has to be calibrated by choosing one cross section (the Compton scattering) to fix the renormalised parameters $\Frac[e/m]$. Only after this calibration one can predict other cross sections and validate or falsify the model.

Gravitational theories are not different even in the classical regime. They have parameters to be calibrated by choosing some conventional observations.
What we are doing is choosing SNIa to fix $\ga$, Mercury's precession to fix $\al$. Further investigation must be devoted to fix $\be$ (e.g.\ by using elements formation).
Only at that point, even when a model has survived calibration, one can use the calibrated model to predict expected values (e.g.\ power spectrum) to falsify the theory on the basis of observations.

In other words, that is the way to go in modified gravity. We are adding parameters (e.g.\ in $f(\cR)$-theories we have the (potentially infinite) parameters which fix the function $f(\cR)$). In order to get stuck to finitely many parameters, we consider a family of functions at a time. Then we need extra experiments to calibrate the theory (in standard GR we only have $G$ which in fact can be fixed by lab experiments) before we are allowed to say we defined a model.
The more parameters we add the harder the calibration, which is what is fair to pay for an extended model of gravitation.

All other heuristic arguments about validity of a theory are based on physical intuition which is often model dependent and moreover it has been developed in standard GR and sometimes uncritically applied to different gravitational theories. We provided above a number of such arguments which eventually have been shown to be inconsistent.
Being stuck to observation is the only robust way in gravitational theories (although probably in general) to really falsify a theory. 

If we take this approach seriously, the current situation is almost desperate for modified gravity models. Let us close by remark and stress that here and in the series of investigation to come we are still trying to falsify a specific family  (\ref{fR})) of Palatini $f(\cR)$-theories. We still have no clue at all about how to do it for a generic Palatini $f(\cR)$-theory, which depends on potentially infinitely many parameters, less then ever for a generic modified gravity theory. However, this is what needs to be done.

\section*{Acknowledgements}

This article is based upon work from COST Action (CA15117 CANTATA), supported by COST (European Cooperation in Science and Technology).
We acknowledge  the contribution of INFN (IS-QGSKY and IS-Euclid), the local research project {\it Metodi Geometrici in Fisica Matematica e Applicazioni (2019)} of Dipartimento di Matematica of University of Torino (Italy). 
SC is supported by the Italian Ministry of Education, University and Research (MIUR) through Rita Levi Montalcini project `{prometheus} -- Probing and Relating Observables with Multi-wavelength Experiments To Help Enlightening the Universe's Structure', and by the `Departments of Excellence 2018-2022' Grant awarded by MIUR (L.~232/2016).
This paper is also supported by INdAM-GNFM.

%
%
%

%

\begin{thebibliography}{}
%
%

\bibitem{Virgo}{The LIGO -Virgo Collaborations,
{\it GW170817: Observation of Gravitational Waves from a Binary Neutron Star Inspiral},
Phys. Rev. Lett. 119 161101 (2017);  \hfill\break
	{\tt arXiv:1710.05832 [gr-qc]}}


\bibitem{ClusterDM1}{Zwicky, F., 
{\it Die Rotverschiebung von extragalaktischen Nebeln}, 
Helvetica Physica Acta, 6: (1933) 110--127}

\bibitem{ClusterDM2}{Zwicky, F., 
{\it On the Masses of Nebulae and of Clusters of Nebulae}, 
Astrophysical Journal, 86: (1937) 217}

\bibitem{GalaxyDM1}{K.C. Freeman,  
{\it On the Disks of Spiral and S0 Galaxies},
Astrophysical Journal 160, 811 (1970)}

\bibitem{GalaxyDM2}{V.C. Rubin,  J.W.K. Ford,
{\it  Rotation of the Andromeda Nebula from a Spectroscopic Survey of Emission Regions},
The Astrophysical Journal 159, 379 (1970).
}

\bibitem{Lensing}{
V. Trimble,   
{\it Existence and nature of dark matter in the universe}. 
Annual Review of Astronomy and Astrophysics. 25: (1987). 425--472.
}

\bibitem{DarkSectiorSupernovae}{A.G. Riess,  et al.,
{\it Observational Evidence from Supernovae for an Accelerating Universe and a Cosmological Constant},
In: AJ 116, (1998) 1009--1038;  \hfill\break
{\tt arXiv:astro-ph/9805201}}


\bibitem{MOND}{
M. Milgrom, 
{\it A modification of the Newtonian dynamics as a possible alternative to the hidden mass hypothesis},
Astrophys. J. 270 (1983) 365.
}

\bibitem{CG1}{
P. D. Mannheim, D. Kazanas, 
{\it Exact Vacuum Solution to Conformal Weyl Gravity and Galactic Rotation Curves}, 
Astrophys.J. 342 (1989) 635--638.}

\bibitem{CG2}{M. Campigotto, L. Fatibene, 
{\it Generally Covariant vs. Gauge Structure for Conformal Field Theories}, 
Annals Phys. 362 (2015) 521?528; \hfill\break
{\tt arXiv:1506.06071}
}

\bibitem{CG3}{
R. Jackiw, S.-Y. Pi, 
{\it Fake Conformal Symmetry in Conformal Cosmological Models};
{\tt arXiv:1407.8545}
}

\bibitem{Od1}{
S.Nojiri, S.D. Odintsov,
{\it Modified gravity with negative and positive powers of the curvature: Unification of the inflation and of the cosmic acceleration},
Phys.Rev. D68 (2003) 123512; \hfill\break
{\tt hep-th/0307288}}

\bibitem{Od2}{
S.Nojiri, S.D. Odintsov,
{\it Introduction to modified gravity and gravitational alternative for dark energy},
in: {\it eConf C0602061 (2006) 06}, Int.J.Geom.Meth.Mod.Phys. 4 (2007) 115-146; \hfill\break
{\tt hep-th/0601213}}

\bibitem{Od3}{
S.Nojiri, S.D. Odintsov,
{\it Unified cosmic history in modified gravity: from F(R) theory to Lorentz non-invariant models},
Phys.Rept. 505 (2011) 59-144; \hfill\break
{\tt arXiv:1011.0544}}



\bibitem{ETC}{Salvatore Capozziello, Mariafelicia F. De Laurentis, Lorenzo Fatibene, Marco Ferraris and Simon Garruto,
{\it Extended Cosmologies},
SIGMA 12 (2016), 006, 16 pages;  \hfill\break
{\tt arXiv:1509.08008}
}


\bibitem{Magnano}{G. Magnano, M. Ferraris, M. Francaviglia, 
{\it Nonlinear gravitational Lagrangians},
Gen.Rel.Grav. {\bf 19}(5), 1987, 465-479}


\bibitem{BD}{C.Brans,  R. H.Dicke, 
{\it Mach's Principle and a Relativistic Theory of Gravitation},
Phys.~Rev.~124(3) (1961) 925--935}

 


\bibitem{reviewPalatinif(R)}{S.Capozziello, M. De Laurentis (2015), 
{\it $F(R)$ theories of gravitation},
Scholarpedia, 10(2):31422.}


\bibitem{Borowiec}{ A.Borowiec, M.Kamionka, A.Kurek, Marek Szyd\l owski,
    {\it Cosmic acceleration from modified gravity with Palatini formalism}, \hfill\break
    	{\tt arXiv:1109.3420 [gr-qc]}
}


\bibitem{Olmo2011}{G.J. Olmo,
     {\it Palatini Approach to Modified Gravity: f(R) Theories and Beyond}, \hfill\break
     	{\tt arXiv:1101.3864 [gr-qc]}
}

\bibitem{OlmoNostro}{L.Fatibene and M.Francaviglia,
{\it Extended Theories of Gravitation and the Curvature of the Universe -- Do We Really Need Dark Matter?}
in : Open Questions in Cosmology, Edited by Gonzalo J. Olmo, Intech (2012), ISBN 978-953-51-0880-1; DOI: 10.5772/52041}





\bibitem{no-go}{E.Barausse, Thomas P.Sotiriou, J.C.Miller, 
{\it A no-go theorem for polytropic spheres in Palatini $f(R)$ gravity},
DOI: 10.1088/0264-9381/25/6/062001\ (4th March 2008)}


\bibitem{Olmo}{G.J.Olmo, 
{\it Re-examination of polytropic spheres in Palatini $f(R)$ gravity},
DOI: 10.1103/PhysRevD.78.104026\ (20th October 2008)}

\bibitem{Mana}{A. Mana, L.Fatibene, M.Ferraris
{\it A further study on Palatini $f(\cR)$-theories for polytropic stars},
JCAP 1510 (2015) 040 (2015-10-16) DOI: 10.1088/1475-7516/2015/10/040;  \hfill\break
{\tt arXiv:1505.06575}
}

\bibitem{Wojnar}
{A. Wojnar,
{\it On stability of a neutron star system in Palatini gravity},
Eur. Phys. J. C (2018) 78: 421;  \hfill\break
{\tt arXiv:1712.01943 [gr-qc]}
}



 \bibitem{MathEquivalence}{ L. Fatibene, M.Francaviglia, 
{\it Mathematical Equivalence versus Physical Equivalence between Extended Theories of Gravitation},
Int. J. Geom. Methods Mod. Phys. 11(1), 1450008 (2014);  \hfill\break
{\tt arXiv:1302.2938 [gr-qc]}
}




\bibitem{ETG}{L.Fatibene, S.Garruto,
{\it Extended Gravity},
Int. J. Geom. Methods Mod. Phys., 11, 1460018 (2014);  \hfill\break
{\tt arXiv:1403.7036 [gr-qc]}       }

\bibitem{Conservation}{L.Fatibene, M.Francaviglia,
{\it Fluids in Weyl Geometries},
IJGMMP, {\bf 09}(2), 1260003 (2012);\hfill\break
{\tt arXiv:1109.4115 [math-ph]}
}


\bibitem{RovelliObs}{C. Rovelli,
{\it What is observable in classical and quantum gravity?},
Class.~Quantum Grav.{8} (1991) 297.
}

\bibitem{book2}{L.Fatibene,
{\it Relativistic theories, gravitational theories and General Ralativity},
in preparation, draft version 1.0.1.\\
{\tt https://sites.google.com/site/lorenzofatibene/my-links/book-version-1-0-1}
}

\bibitem{Distances}{S.Capozziello, A.Chiappini, L.Fatibene,  A. Orizzonte,
{\it The generally covariant meaning of space distances},
EPJC xx
}

\bibitem{Perlick}{V.Perlick,
{\it Characterization of Standard Clocks by Means of Light Rays and Freely Falling Particles},
Gen. Rel. Grav. {\bf 19}(11), (1987) 1059-1073
}

\bibitem{Pinto1}{P.Pinto, L.Del Vecchio, L.Fatibene, M.Ferraris,
{\it Extended Cosmology in Palatini $f(\cR)$-theories},
(submitted);\hfill\break
{\tt arXiv:1807.00397 [gr-qc]}
}


\bibitem{HubbleDrift}{
L.Del Vecchio, L.Fatibene, S.Capozziello, M.Ferraris, P.Pinto, S.Camera,
{\it Hubble drift in Palatini $f(\cR)$-theories},
Eur. Phys. J. Plus 134(5) (2019);\hfill\break
arXiv:1810.10754 [gr-qc]
}

\bibitem{Weinberg}{S. Weinberg,
{\it Gravitation and Cosmology: Principles and Applications of the General Theory of Relativity}
Wiley, New York (a.o.) (1972). XXVIII, 657 S. : graph. Darst.. ISBN: 0-471-92567-5.}

\bibitem{Universality}{A. Borowiec, M. Ferraris, M. Francaviglia, I. Volovich,
{\it Universality of Einstein Equations for the Ricci Squared Lagrangians},
Class. Quantum Grav. 15, 43-55, 1998}


\bibitem{PlanckExp}{Planck Collaboration,
{\it Planck 2015 results. XIII. Cosmological parameters},
	A\&A 594, A13 (2016);  \hfill\break
{\tt arXiv:1502.01589 [astro-ph.CO]}
}

\bibitem{CMB}{F. Melchiorri, B.O. Melchiorri,  L. Pietranera,  B.O. Melchiorri,  
{\it Fluctuations in the microwave background at intermediate angular scales}, 
 The Astrophysical Journal. 250: L1, (1981).}

\bibitem{WMap}{E. Komatsu et al.,
{\it Five-Year Wilkinson Microwave Anisotropy Probe (WMAP) Observations: Cosmological Interpretation},
 2009 ApJS 180 330; \hfill\break 
{\tt  arXiv:0803.0547 [astro-ph]}}


 \bibitem{Cassini1}{L.Perivolaropoulos, 
 {\it PPN parameter $\ensuremath{\gamma}$ and Solar System constraints of massive Brans-Dicke theories},
 Phys.~Rev.~D 81(4) 047501-4 (2010);\hfill\break
 arXiv:0911.3401 [gr-qc]
 }
 
 \bibitem{Cassini2}{A.Hees, A.F\"uzfa,
 {\it Combined cosmological and Solar System constraints on chameleon mechanism},
 Phys.~Rev.~D 85(10) 103005-21 (2012);\hfill\break
 arXiv:1111.4784 [gr-qc]
 }


\bibitem{Bhadra}{A. Bhadra,
{\it Brans-Dicke theory: Jordan vs Einstein Frame},
Phys.Rev.D74:014016,2006;\goodbreak
hep-ex/0605109
}


\bibitem{Kozyrev}{
Sergey Kozyrev,
{\it Exact Vacuum Solutions of Jordan, Brans-Dicke Field Equations}; \goodbreak
gr-qc/0512020
}






\bibitem{EPS}{J.Ehlers, F.A.E.Pirani, A.Schild, 
    {\it The Geometry of Free Fall and Light Propagation}, 
    in: General Relativity, ed. L.O.`Raifeartaigh (Clarendon, Oxford, 1972).}
    
\bibitem{EPSNostro}{ M. Di Mauro, L. Fatibene, M.Ferraris, M.Francaviglia, 
{\it Further Extended Theories of Gravitation: Part I }
Int. J. Geom. Methods Mod. Phys. Volume: 7, Issue: 5 (2010), pp. 887-898;  \hfill\break
{\tt gr-qc/0911.2841}}


\bibitem{Euclid}{R. Scaramella et al.,
{\it Euclid space mission: a cosmological challenge for the next 15 years},
in {\it Proceedings IAU Symposium} No. 306, 2014, "Statistical Challenges in 21st Century Cosmology", A.F. Heavens, J.-L. Starck \& A. Krone-Martins, eds;
 \hfill\break
{\tt  arXiv:1501.04908}
 }
 
\bibitem{HD}{
 P.-S. Corasaniti, D.Huterer, A. Melchiorri,
{\it  Exploring the Dark Energy Redshift Desert with the Sandage-Loeb Test},
Phys.Rev.D75:062001,2007;  \hfill\break
{\tt arXiv:astro-ph/0701433}
}













\end{thebibliography}
%

\end{document}